\documentstyle[aps, epsf, prl, preprint, tighten]{revtex}
\begin{document}
\draft
%
\hoffset= -2.5mm

\title{Classical to quantum crossover in high-current noise
       of one-dimensional ballistic wires}

\author{Frederick Green${}^{1,2}$
and Mukunda P. Das${}^2$}
\address{
${}^1$Centre for Quantum Computer Technology,
School of Physics, University of New South Wales,
NSW 2052, Australia.}
\address{
${}^2$Department of Theoretical Physics,
Research School of Physical Sciences and Engineering,
Institute of Advanced Studies,\\
The Australian National University, ACT 0200, Australia.}

\maketitle

\begin{abstract}
Microscopic current fluctuations are inseparable from conductance.
We give an integral account of both quantized conductance and
nonequilibrium thermal noise in one-dimensional
ballistic wires. Our high-current noise theory
opens a very different window on such systems. 
Central to the role of nonequilibrium
ballistic noise is its direct and robust dependence
on the statistics of carriers. For, with increasing density,
they undergo a marked crossover from classical to strongly
degenerate behavior. This is singularly evident
where the two-probe conductance shows quantized steps:
namely, at the discrete subband-energy thresholds.
There the excess thermal noise of field-excited
ballistic electrons displays sharp and large peaks,
invariably larger than shot noise.
Most significant is the nonequilibrium peaks'
high sensitivity to inelastic relaxation
within the open system.
Through that sensitivity, high-current noise provides
unique clues to the origin of quantized contact resistance
and its evolution towards normal diffusive conduction.
\end{abstract}

\section{Introduction}

In this work we address a major, yet largely neglected, issue of
mesoscopic transport:
the nature of high-field thermal fluctuations and
of what they reveal about transport dynamics.
Noise and conductance in electronic transport are intertwined.
Knowledge of one illuminates the other. Nowhere is their
symbiosis clearer, or more critical to physical understanding,
than in mesoscopic conduction.

Carrier fluctuations, measured as noise, change dramatically in
the physical setting of ballistic transport
\cite{blbu}.
Noise yields insights into the innermost
aspects of carrier motion that cannot be gained from conductance
alone. Noise at {\em high currents} is the most revealing.
Little is known about high-field mesoscopic noise, despite the ease
with which even modest driving potentials can push
a mesoscopic system beyond linear response
\cite{gdi}.
It is acknowledged that this unfamiliar
but important region must be opened up and mapped
\cite{blbu}.

We preface our study with a brief recollection of ballistic transport.
So-called mesoscopic conductors have sizes
comparable to the scale for electron scattering.
Such systems no longer exhibit bulk Ohmic behavior.
This was strikingly demonstrated in the recent
experiments of de Picciotto {\em et al.}
\cite{depic}.
The resistance of a ballistic, almost
one-dimensional (1D) quantum wire was probed, noninvasively,
at two inner contacts well separated from the
large diffusive leads acting as current source and drain.
Across the inner probes and at finite current,
the wire's voltage drop, and hence the intrinsic resistance,
were deduced to be essentially zero.

Reference
\onlinecite{depic}
also reported two-probe current-voltage data that
included the source and drain in series with the 1D channel.
In their well characterized 1D wire they found good agreement
with the standard Landauer-B\"uttiker (LB) prediction
of quantized jumps in the two-terminal conductance,
as a function of carrier density
\cite{ferry}.

At this point we come upon the core issue in ballistic transport.
How can the finite conductance (albeit with sharp steps)
be reconciled with the property of totally resistance-free
transport within the central ballistic segment?

The LB theory accounts for this coexistence by simple kinematics.
A 1D wire admits a sparse set of discrete subband states
for electron transmission, while its attached leads have
a much richer quasicontinuous density of such states. As well,
they are collisionally coupled to a rich phonon spectrum.
Bottlenecks arise as carriers try to funnel into and fan out of the
small set of 1D states. That mismatch destroys perfect conduction
and manifests as the ``contact resistance'' of the leads.

According to the LB account, the contact resistance is quantized
because its bottleneck is regulated {\em solely} by the occupancy of
the quantized ballistic states.
Each time the electrons' Fermi level accesses an additional higher
1D subband, the ideal two-probe conductance is shown to
jump by $e^2/\pi\hbar$
\cite{ferry}.

In the Landauer-B\"uttiker picture, therefore, the two-probe
conductance appears to be {\em free} of any influence from extraneous
parameters such as the asymptotic density of states and
scattering rates for the macroscopic leads;
those can enter only implicitly in the transport
\cite{iml}.
One must suppose that the physics of relaxation in
the carrier reservoirs remains
concealed -- in a nonspecific fashion -- within
each lead's equilibrium state
as it fixes the boundary conditions for the sample
\cite{iml}.

\section{A Physical Dilemma}

We make an important observation. If a 1D wire is imperfectly
transmissive (finite resistance), LB predicts
departures from the ideal staircase for two-probe conductance.
On the other hand, if one knows for sure that the 1D wire
transmits {\em perfectly} (zero resistance) LB is bound to say,
unequivocally, that
the two-probe conductance must scale strictly universally.
The published two-probe results of Ref. \onlinecite{depic}
record a somewhat imperfect ``universal'' conductance,
yet the four-probe results record perfect transmission.
We will return to this apparent dilemma.

Here noise enters as a wholly different kind of probe.
Beside the linear LB theory for conductance,
there is a corresponding theory for low-field mesoscopic
fluctuations (in the strictly linear regime, these should
always be be tightly circumscribed
by the fluctuation-dissipation theorem).
The most authoritative survey of LB noise theory to date
is Blanter and B\"uttiker's
\cite{blbu}.

Inherently nonlinear mesoscopic noise
is out of reach to the LB approach,
which can treat only voltages much below the Fermi energy
\cite{blbu}.
We now pose our key question: What can those little-known
fluctuations reveal about the wire-lead interaction and the
contact resistance?

We predict the surprising existence of large peaks in the
thermal, or hot-electron, noise of a ballistic system.
These noise structures are {\em much larger than shot noise}
which also peaks at the stepwise two-probe conductance transitions.
For high-field thermal noise, more can be said.
Its maxima carry unique quantitative information
tracking the interplay of
inelastic and elastic scattering in the leads.
Nonlinear fluctuations, governed by
inelasticity, are simply inaccessible to
elastic-only, low-field descriptions.
For our part, we build upon existing nonequilibrium kinetics
\cite{gdi,gd2}.

In the following Section we describe our kinetic model for
the conductance of a 1D ballistic wire fully open to
an incoherent dissipative environment.
In Sec. 4 we extend the kinetic description to
the nonequilibrium current noise in the system,
and demonstrate the rich behavior of its spectral density.
Section 5 has our numerical results, confirming that
hot ballistic electrons, in their transition from a classical to a 
degenerate regime, yield dramatic thermal-fluctuation peaks.
Finally, in Sec. 6 we sum up; our predictions for
nonequilibrium ballistic noise have direct implications
for corroborating the recent seminal four-probe data
\cite{depic}.

\section{Kinetic Description}

Figure 1 depicts our model system. An ideal
1D wire of length $L$ is connected to two large reservoirs
which are always at equilibrium; the identical arrangement of
Landauer-B\"uttiker, as for Ref.
\onlinecite{depic}.
Elastic (nondissipative) and inelastic (dissipative) collisions
occur within the leads. A kinetic description will cover both
{\em wire and leads}. Inelastic and elastic mean free paths
(MFPs) cannot be shorter than $L$.


We stress that it is utterly essential to incorporate,
within the microscopic equations of motion, at least
the leading-order effects of dissipative inelastic collisions in the leads.
It is the kinetics of dissipative relaxation {\em alone} that
secures the energetic stability of transport over the system
\cite{iml},
by establishing steady state.

There is a second crucial consequence of inelastic
relaxation in the leads: the total loss of phase memory for
carriers crossing the ballistic region of the system.
From the ballistic carriers' perspective, they no longer
bear the detailed signature of
their collision history, which thus becomes Markovian.

The only kinetic parameters that survive phase breaking
are (a) the effective scale of the scattering mean free paths
(matching the ballistic length),
and (b) the size of the driving field that impels
carriers across the collision-free segment.
It is precisely through phase breaking that the ballistic
system becomes insensitive to the particulars of
geometry and material structure
of its asymptotic leads.
On that basis, one can make a semi-classical analysis
of the ballistic kinetics.

Our Fermi-liquid perspective is very closely related to that of 
Kamenev and Kohn
\cite{kk}.
We generalize their coherent, closed-circuit, linear-response approach
to deal with a phase-broken, open system driven far from
equilibrium.
We guarantee the open system's all-important {\em gauge invariance}
by inserting an explicit flux $I/e$ of electrons
at the source, and its matched sink $-I/e$ at the drain
(equivalently, a hole flux of strength $I/e$).

Formally, the direct inclusion of boundary flux sources
is strictly mandated by
global -- as distinct from local -- charge
and current conservation in open conductors,
with nontrivial electromotive forces (EMFs)
acting from outside
\cite{sols}.
Physically, the flux-source regions at the lead--wire boundaries
fulfill a unique, and absolutely necessary, dynamical role. They are
the active sites for the nonequilibrium scattering processes that
directly determine the relaxation in the leads.

First we recover the
full Landauer-B\"uttiker conductance by our kinetic analysis.
Consider any single, occupied 1D subband.
From the viewpoint of their time-stationary distribution,
incoherence in the leads means that carriers inside the finite wire
will perceive a homogeneous physical environment, not only within the
wire proper (internal homogeneity is also assumed in the LB description)
but at a far longer range.
They will behave as though the physical wire were
embedded, seamlessly, in a very long effective 1D host. This
effective host conductor, ``standing in'' for the leads,
is itself nonideal {\em on a scale to match the ballistic length} $L$.

The boundaries and field sources
are as if at infinity, even though the actual
current injector and extractor can be adjacent to the wire.
Explicit carrier relaxation in the boundary neighborhoods,
in concert with the EMF (Landauer's resistivity dipole),
asymptotically ``prepares'' the
nonequilibrium {\em and uniform} steady state of the carriers
crossing the inner ballistic region.
The size $V$ of the EMF quantifies
the injected carriers' {\em self-consistent} adjustment
to relaxation inside the embedding host (the leads).
The collision-mediated relation between
$V$ and $I$ follows
\cite{rl57}.

Let the steady-state electron distribution in the wire be
$f_k$ for states $k$ within our 1D subband (spin label implied).
The collision-time form for
the kinetic equation,
for EMF field $-E$ in the source-to-drain
direction, is
\cite{stanton,gdi}

\begin{equation}
{eE\over \hbar}
{\partial f_k\over \partial k}
=
-{1\over \tau_{\rm in}(\varepsilon_k)}
{\left( f_k -
{{\langle \tau_{\rm in}^{-1} f \rangle}\over 
 {\langle \tau_{\rm in}^{-1} f^{\rm eq} \rangle}}
f^{\rm eq}_k
\right)}
-{1\over \tau_{\rm el}(\varepsilon_k)}
{ {f_k - f_{-k}}\over 2 }.
\label{e1}
\end{equation}

\noindent
Here $\tau_{\rm in}(\varepsilon_k)$
and $\tau_{\rm el}(\varepsilon_k)$ are the
inelastic and elastic scattering times,
in general energy-dependent.
The leading, inelastic
collision term has a restoring contribution
proportional, in the general case, to the expectation

\[
{\langle \tau_{\rm in}^{-1} f(t) \rangle}
\equiv \int^{\infty}_{-\infty}
{2dk\over 2\pi} \tau_{\rm in}^{-1}(\varepsilon_k) f_k(t).
\]

\noindent
The inelastic collision term respects continuity
and (local) gauge invariance
\cite{quinn,stanton}
dynamically and, as in Eq. (\ref{e1}), statically. 
As usual, the underlying equilibrium distribution is
$f^{\rm eq}_k = 1/{\{ 1 +
\exp[(\varepsilon_k + \varepsilon_i - \mu)/k_{\rm B}T] \}}$
where $\varepsilon_i$ is the band-threshold energy. 
{\em Only one chemical potential}, $\mu$, {\em enters the problem}
\cite{gdi,kk,gd2}.
Finally, the elastic collision term restores symmetry.

A microscopic formulation such as Eq. (\ref{e1}) does not
need to segregate left- and right-moving carriers
for exclusive treatment
\cite{iml}.
Nor, as Kamenev and Kohn have rigorously shown
\cite{kk},
does such a formalism need the special creation of
chemical potentials exclusive to different movers
\cite{iml}.

Let us now specialize, just as in Landauer-B\"uttiker,
to energy-independent scattering rates.
Stanton
\cite{stanton}
has solved Eq. (\ref{e1}) analytically
with a particular linear transformation:

\begin{equation}
{\widehat f}_k
\equiv
  {1\over 2}{\left( 1 + \sqrt{\tau_{\rm in}\over \tau} \right)} f_k
+ {1\over 2}{\left( 1 - \sqrt{\tau_{\rm in}\over \tau} \right)} f_{-k}
\label{e2}
\end{equation}

\noindent
where $\tau^{-1} = \tau_{\rm in}^{-1} + \tau_{\rm el}^{-1}$ is
recognizable as the Matthiessen relaxation rate.
The composite function ${\widehat f}$
satisfies a simpler form of Eq. (\ref{e1})
in which $\sqrt{\tau\tau_{\rm in}}$ appears in
place of $\tau_{\rm in}$ and there is no elastic collision term.
Its exact solution is
\cite{stanton}

\begin{equation}
{\widehat f}_k
= \lambda \int^k_{-\infty} dk' e^{-\lambda(k - k')} f^{\rm eq}_{k'},
\label{e3}
\end{equation}

\noindent
where $\lambda = \hbar/(eE\sqrt{\tau\tau_{\rm in}})$.
The physical solution is easily recovered, as is
the expectation value of the current:

\begin{equation}
I = {\langle ev_k f_k \rangle}
= {e\hbar\over m^*}{\langle k f_k \rangle}
= {e\hbar\over m^*}\sqrt{\tau\over \tau_{\rm in}}
{\langle k {\widehat f}_k \rangle}
\label{e6a}
\end{equation}

\noindent
for a parabolic subband with effective mass $m^*$.
Integration at subband density $n$ yields the time-honored result

\begin{equation}
I = \sqrt{\tau\over \tau_{\rm in}} {n e\hbar\over m^*\lambda}
= {ne^2\tau\over m^*}E.
\label{e6b}
\end{equation}

Near equilibrium the ballistic hypothesis applies.
The dominant mean free paths
in the effective 1D host conductor, namely the Fermi ones
${v_{\rm F}\tau_{\rm in}}$ and ${v_{\rm F}\tau_{\rm el}}$,
will each span $L$, the ideal collision-free length of the
uniform sample. Let us therefore equate each MFP to $L$;
then $\tau = L/2v_{\rm F}$.
The uniform-state solution Eq. (\ref{e6b}) gives, on writing the
subband density as $n = 2m^*v_{\rm F}/\pi \hbar$,

\begin{equation}
I = {2m^*v_{\rm F}\over \pi\hbar} {e^2 EL\over 2m^*v_{\rm F}}
= {e^2\over \pi\hbar} V
\label{e6c}
\end{equation}


\noindent
in which $V = EL$ is the external EMF.
Equation (\ref{e6c}) exhibits precisely
the Landauer-B\"uttiker conductance.

This result complements the Kamenev-Kohn approach,
which derives Eq. (\ref{e6c}) by applying {\em orthodox}
Fermi-liquid principles (via microscopic Kubo and
quantum-transmission theories) in the context of a
closed, nondissipative, phase-coherent mesoscopic circuit
\cite{kk}.
Equally, the present derivation builds upon
standard Fermi-liquid microscopics and applies it
(via kinetic theory) to the open, dissipative, phase-breaking
context of mesoscopic circuits in the laboratory
\cite{gdi,gd2}.

\section{Nonequilibrium Ballistic Fluctuations}

Our open-system kinetic theory
for ballistic transport clearly leads to
the Landauer-B\"uttiker quantized conductance.
Now we obtain the nonequilibrium hot-electron noise,
which the linear-response LB picture cannot attain.
For this we need the retarded, space-time dependent Green function
$R_{kk'}(x,x'; t - t') = \theta(t - t')
\delta f_k(x,t)/\delta f_{k'}(x',t')$ for the
dynamical form of Eq. (\ref{e1}).
The standard, manifestly gauge-invariant
kinetic equation for the dynamic response $R$ is
\cite{stanton,gdi,ggk,sw}

\begin{eqnarray}
{\Biggl[
{\partial\over \partial t}
+ v_k{\partial\over \partial x}
+ {eE\over \hbar} {\partial\over \partial k}
\Biggr]}
R_{kk'}(x,x'; t - t')
=&&
2\pi\delta(k - k') \delta(x - x') \delta(t - t')
\cr
{\left. \right.} \cr
&&
-{1\over \tau_{\rm in}}
{\left(
R_{kk'} - {\langle R_{k''k'} \rangle}'' {f^{\rm eq}_k\over n}
\right)}
\cr
{\left. \right.} \cr
&&
-{1\over 2\tau_{\rm el}}
(R_{kk'} - R_{-k,k'}).
\label{e7}
\end{eqnarray}

The first step in solving $R_{kk'}(x,x'; t - t')$
is to Fourier transform it
to ${\cal R}_{kk'}(q,q';\omega)$ in the
momentum-frequency domain $(q,q';\omega)$.
Next, a frequency-dependent continuation of Stanton's
transformation for ${\widehat f}$,
Eq. (\ref{e2}), leads one to a composite Green function
${\widehat {\cal R}}_{kk'}(q,q';\omega)$. That object,
a well formed linear combination of matrix elements
of the physical response ${\cal R}_{kk'}(q,q';\omega)$,
has an equation of motion more readily solved
than Eq. (\ref{e7}). The last major step is to
isolate the purely transient component of 
${\widehat {\cal R}}$. This is
embodied in the correlated propagator
\cite{gdi,ggk}

\begin{equation}
{\widehat {\cal C}}_{kk'}(q,q';\omega)
\equiv
{\widehat {\cal R}}_{kk'}(q,q';\omega)
- j_0(qL/2)
{\langle {{\widehat {\cal R}}_{k''k'}(0,q';\omega) \rangle}}''
{{\widehat f}_k\over n},
\label{e7.1}
\end{equation}

\noindent
whose second right-hand term represents the adiabatic
(low-frequency dominant) contribution to
${\widehat {\cal R}}$, now subtracted
\footnote{
The spherical Bessel function $j_0(qL/2)$ is the
transform of the step function $\theta(L/2 - |x|)$,
delimiting the finite ballistic wire.
}.

The current-current spectral density,
taken over all carriers in the sample,
can now be regained. It is a linear superposition
of convolutions of the correlation
$(\pm v_k)(\pm v_{k'}){\widehat {\cal C}}_{\pm k,\pm k'}$,
evaluated for all sign choices, with the {\em steady-state}
mean-square fluctuation distribution

\begin{equation}
\Delta {\widehat f}_{k'}
= \lambda \int^{k'}_{-\infty}
dk'' e^{-\lambda(k' - k'')} k_{\rm B}T
{\partial f^{\rm eq}_{k''}\over \partial \mu}.
\label{e8}
\end{equation}

\noindent
With the same linear transformation used in Eq. (\ref{e2}),
this uniquely maps the static electron-hole pair correlation
at equilibrium (the last derivative factor in the integrand
of Eq. (\ref{e8})) to its fully nonequilibrium counterpart
\cite{gdi}.

In the dynamics
of a spontaneous thermal excursion of the steady state,
Eq. (\ref{e8}) fixes the mean initial strength
of the background electron-hole fluctuation $\Delta {\widehat f}$,
perturbing the nonequilibrium system at any given instant.
From Eq. (\ref{e7.1}),
${\widehat {\cal C}}$ then provides its subsequent relaxation.

The flux autocorrelation function has the raw form
${\langle {\langle v {\widehat C} v'
\Delta {\widehat f'} \rangle} \rangle}'$.
After some tedious but straightforward linear algebra,
the original quantities $C_{kk'}(q,q';\omega)$ and
$\Delta f_{k'}$ are reconstructed and, with them, we
finally obtain the complete and explicit physical form
${\langle {\langle v C v' \Delta f' \rangle} \rangle}'$.
This defines the thermal-noise spectral density in the
ballistic wire 
\cite{gdi,gd2,ggk},
specializing now to the $i$th subband:

\begin{equation}
{\cal S}_i(q,q';\omega) \equiv
4 {\Re {\Bigl\{
{\langle {\langle
(-ev/L) C_i(q,q';\omega) (-ev'/L) \Delta f_i'
\rangle} \rangle}'
\Bigr\}} }.
\label{e8.1}
\end{equation}

The long-time average of the hot-electron fluctuations,
integrated over the wire for a parabolic subband at density
$n_i = 2m^*v_{{\rm F}i}/\pi\hbar$,
comes from the long-wavelength static form of
Eq. (\ref{e8.1}):

\begin{eqnarray}
{\cal S}^{\rm xs}_i(V)
=&& {\cal S}_i(0,0;0) - 4G_ik_{\rm B}T
\cr
{\left. \right.} \cr
=&&
4G_i k_{\rm B}T
{{\partial \ln n_i}\over {\partial \mu}}
{e^2V^2\over 2m^* L^2}
{\Bigl(\tau_{{\rm in};i}^2 + 2\tau_i \tau_{{\rm in};i} - \tau_i^2\Bigr)};
\label{e9}
\end{eqnarray}

\noindent
note that the (dissipative) Johnson-Nyquist term
$4G_ik_{\rm B}T$ is removed.
The subband conductance is $G_i = (2v_{{\rm F}i}\tau_i/L)G_0$
with $G_0 = e^2/\pi\hbar$.
We do not necessarily assume, as for Eq. (\ref{e6c}),
collision times $\tau_{{\rm el};i}$ and  $\tau_{{\rm in};i}$
equal to the transit time $L/v_{{\rm F}i}$
and predicated on a potentially singular density
dependence of the relaxation rates.

Equation (\ref{e9}) is clearly sensitive to the
ratio $\tau_{{\rm in};i}/\tau_{{\rm el};i}$. It
probes the dynamics imposed by the leads
{\em upon each individual subband}.
It is our foremost outcome.
 
There are three points more.

\smallskip
\noindent
(\i) The excess thermal noise is nonlinear in
the EMF. Although, like shot noise, it is
nondissipative
\cite{gdi,dtg},
Eq. (\ref{e9}) {\em does not describe shot noise}
\cite{blbu}.
We emphasize that ${\cal S}^{\rm xs}$
is an entirely separate and experimentally
distinguishable effect.

\smallskip
\noindent
(\i\i) For strong inelastic
scattering ${\cal S}^{\rm xs}_i/G_i$ vanishes.
At weak inelasticity it {\em diverges};
as $\tau_{{\rm in};i} \to \infty$, carriers
have no way to shed excess energy and the hot-electron
distribution broadens without limit.
Fluctuations are all. Thus,
transport models based solely on {\em elastic}
relaxation have, at best, a marginal thermodynamic stability.

\smallskip
\noindent
(\i\i\i) At strong degeneracy ${\cal S}^{\rm xs}_i$ scales as
$k_{\rm B}T{\partial \ln n_i/\partial \mu}
= k_{\rm B}T/2(\mu - \varepsilon_i)$. This is
a feature generic to the dense nonequilibrium electron gas
\cite{gdi,gd2}.
For classical electrons, the same factor goes to unity,
and classical excess noise is then independent of $T$.

\section{Numerical Results}

In the light of (\i\i\i) above, the nonequilibrium behavior of
Eq. (\ref{e9}) at a subband threshold is extremely interesting.
An increase in electron density within each level $i$
takes its ballistic carrier population out of the classical domain
($\mu \ll \varepsilon_i$) and makes it cross over to the
highly degenerate regime ($\mu \gg \varepsilon_i$).
Thus the kinematic $T$-scaling of ${\cal S}^{\rm xs}$
starts to attenuate it strongly while, {\em at the same time},
the factor $G_i$ in Eq. (\ref{e9}) grows from almost nothing
at low subband occupancy to values near $G_0$ at high density.
This pattern repeats itself as each level is filled in succession.

Evidently, classical statistics naturally dominates whenever
$\mu - \varepsilon_i \leq k_{\rm B}T$.
If one insists on suppressing it by imposing a
strict but inappropriate degeneracy there,
the development of ${\cal S}^{\rm xs}$ will echo that
for the Landauer-B\"uttiker conductance model:
$G_i \to \theta(\mu - \varepsilon_i)G_0$.
Then ${\cal S}^{\rm xs}$ behaves pathologically,
and quite unphysically.
Forgoing this unnatural choice we shall adopt
a proper finite-temperature representation, with no
artificially enforced degeneracy at the subband crossings.

The Ansatz in Eq. (\ref{e6c}), whereby
$\tau_i \sim 1/{v_F}_i$ as ${v_F}_i \to 0$, must be
corrected for its spurious divergence when 
the chemical potential falls below the subband edge.
The MFP for {\em elastic} impurity scattering
stays close to $L$ and is insensitive to hot-electron effects,
while the inelastic MFP shortens as phonon emission sets in.
We write the elastic time as $\tau_{{\rm el};i} = L/u_i(\mu)$ for the
well-behaved and characteristic velocity average
$u_i(\mu) \equiv 2{\langle |v| f^{\rm eq} \rangle}_i/n_i$.
Near and below the subband crossing $u_i$ is
the classical, thermal velocity; above, $u_i \to v_{{\rm F}i}$.
We will map the behavior of ${\cal S}^{\rm xs}$
as a functional of the more labile {\em inelastic} time,
which we parametrize as
$\tau_{{\rm in};i} \equiv \zeta_i \tau_{{\rm el};i}$.

Our results for a two-band model are in Fig. 2.
We choose inelastic-to-elastic ratios
$\zeta_1 = 1$ and $\zeta_2 = 1$, 0.8, and 0.6
\footnote{
The corresponding LB transmission factors\cite{blbu} are
given by ${\cal T}_i = G_i/G_0$.
In the degenerate limit this means that
${\cal T}_i \to 2\zeta_i/(1 + \zeta_i)$ so that, for
our chosen values $\zeta_i$, one obtains the respective
LB factors
${\cal T}_1 = 1$ and ${\cal T}_2 = 1$, 0.89, and 0.75.}.
Our total excess noise is
${\cal S}^{\rm xs} = {\cal S}^{\rm xs}_1 + {\cal S}^{\rm xs}_2$,
total conductance is $G = G_1 + G_2$.
Subband edges are $\varepsilon_1 = 5k_{\rm B}T$
and $\varepsilon_2 = 17k_{\rm B}T$.
We keep $V$ constant at $9k_{\rm B}T/e$
in each of three plots of
${\cal S}^{\rm xs}(V)$ as a function of chemical potential.


In ${\cal S}^{\rm xs}$, with increasing $\mu$,
we see dramatic peaks where $G$ has characteristic steps.
The peaks reveal an orderly metamorphosis of 1D hot-electron
fluctuations, from their classical ($T$-independent) regime
at the steps to their degeneracy-suppressed ($T$-scaled)
state at the plateaux. This striking, chameleon-like
transformation bears no relation to shot noise
\cite{blbu}.
Such an effect can be readily probed in the laboratory,
by appropriately biasing a gate voltage that couples
to the electron density in the ballistic channel
\cite{depic,rez}. 

We also see how the thermal-noise maxima change
significantly more with $\zeta_i$ than $G$ itself does;
${\cal S}^{\rm xs}(V)$ is truly a fine marker for
{\em nonuniversal} effects. Experimentally, this could well include
the observation of higher-order phenomena
(lead geometry,  inter-subband transitions
\cite{sw2},
etc.) not covered at our present level of treatment.

It is instructive to continue the study of ${\cal S}^{\rm xs}(V)$
down to low EMFs, comparing it with the well known
Landauer-B\"uttiker ``noise crossover''
formula, which combines all excess thermal noise
into a single entity with shot noise
\cite{blbu}:

\begin{equation}
{\cal S}^{\rm LB}(V) = \sum_i 4G_ik_{\rm B}T
{\left( 1 - {G_i\over G_0} \right)}
{\left[ {eV\over 2k_{\rm B}T}
{\rm coth}{\left( {eV\over 2k_{\rm B}T} \right)} - 1 \right]}.
\label{e12}
\end{equation}

\noindent
The quantitative contrast between ${\cal S}^{\rm xs}$
and the total LB noise is made clear in Fig. 3
for the identical two-band situation of Fig. 2, but now
at $V = 0.9k_{\rm B}T/e$. Even in this low-field regime
we find that ${\cal S}^{\rm xs}$, computed strictly within
an orthodox and fully gauge-invariant kinetic approach,
still dominates the result of the widely held crossover model
\footnote{The Landauer-B\"uttiker noise formula is
{\em not} inherently gauge-invariant. Therefore
its underlying dynamical fluctuations are
not guaranteed to conserve charge.
See Ref. \onlinecite{blbu}, remarks following their Eq. (51).}.


\section{Conclusion: a New Ballistic-Noise Experiment}

We end by recalling the apparent impasse
raised by the de Picciotto {\em et al}. data
\cite{depic},
vis \`a vis the LB theory's categorical and uncompromising
prediction of perfect, universal conductance steps
\cite{blbu,ferry}.
Resolution depends crucially on the physics
of the noninvasive probes. Measurements of
the {\em nonequilibrium} noise would provide a unique
opportunity to reveal how, precisely, that physics acts.

Suppose the noninvasive contacts do access the local potentials.
Had they detected any voltage at all at probe separation $l$ inside
the wire, it would have been the resistivity-dipole potential
$El$. The probes saw nothing.
This implies total neutralization of Landauer's resistivity
dipole within the wire body; $E = 0$. Its canceling counter-fields
must be from dipoles sitting hard by the boundaries.
That makes them highly localized.

The notion of strong additional localized counter-charges,
over and above the natural Hartree displacement
\cite{kk}
that sets up the resistivity dipole in the first place,
is energetically unlikely. Landauer's self-consistent dipole is
the fundamental element of mesoscopic transport
\cite{ferry,rl57}.
Further substructures, at screening scales shorter
than natural, would invite a logical {\em reductio ad absurdum}.

Now suppose instead that the probes couple capacitively
to the 1D wire. At most, they will record variations of
carrier density between their locations. But the wire is uniform,
so there is no variation. Capacitive probes will report no difference.
Thus there is no reason for the resistivity dipole to cancel out.

The conclusion of Ref.
\onlinecite{depic},
that their noninvasive probe arrangement gives genuine and direct
access to internal voltages, may now be checked by
quite different means.
The nonequilibrium {\em thermal} ballistic noise
(totally separate from any considerations of the shot noise
\cite{blbu})
can provide independent experimental verification.

In brief: if the driving field $E$ throughout the ballistic structure
is indeed zero, there will be little excess noise of the kind
we predict. If, on the other hand, $E \neq 0$ then thermal noise
must appear -- and in copious amounts -- at the subband crossings.

Here, the outcome is plain in terms of nonequilibrium noise.
The experimental import of our work is much wider, however.
It covers not just noise in a heterostructure-based device
\cite{depic,rez},
but also fluctuations at the intense fields sustainable
in metallic carbon nanotubes
\cite{kane}.
There, the quantum-confinement energies are huge by comparison
with normal semiconductor-based 1D channels.

Carbon nanotubes provide a far more challenging testing-ground
for transport and fluctuation theories in the extremely high-field
domain. It is an area that is certainly
wide open -- and ripe -- for exciting explorations.

\begin{figure}
FIG. 1.
An ideal, uniform ballistic wire. Its diffusive leads
(S, D) are at equilibrium. A paired source and sink of current
$I$ at the boundaries drive the transport. Local charge clouds
(shaded), induced by the influx and efflux of $I$, set up
the dipole potential $E(I)L$ between D and S.
\label{f1}
\end{figure}

\begin{figure}
FIG. 2.
Left scale: excess thermal noise ${\cal S}^{\rm xs}$
of a ballistic wire at voltage $V = 9k_{\rm B}T/e$,
as a function of chemical potential.
Right scale: two-probe conductance $G$.
At the subband crossing points of $G$, the excess noise peaks.
Noise is high at the crossing points, where
subband electrons are {\em classical}, low at the plateaux
where subband {\em degeneracy} is strong.
${\cal S}^{\rm xs}$, far more than $G$, is sensitive
to the scattering-time ratios
$\zeta_i = \tau_{{\rm in};i}/\tau_{{\rm el};i}$.
Dashed line: ideal LB shot-noise prediction (see Eq. (\ref{e12}))
corresponding to our full line
($\zeta_1 = \zeta_2 = 1$). The predicted
shot noise is much smaller.
\label{f2}
\end{figure}

\begin{figure}
FIG. 3.
Full line: low-field excess thermal noise ${\cal S}^{\rm xs}$
of our two-band ballistic wire, at $V = 0.9k_{\rm B}T/e$.
The scattering-time ratios are ideal: $\zeta_1 = \zeta_2 = 1$.
Dashed line: the corresponding shot-noise prediction
${\cal S}^{\rm LB}$
given by the Landauer-B\"uttiker crossover formula
\cite{blbu}, Eq. (\ref{e12}).
Our standard kinetic-theoretical prediction for the excess noise
dominates the crossover even in the weak-field regime.
\label{f3}
\end{figure}

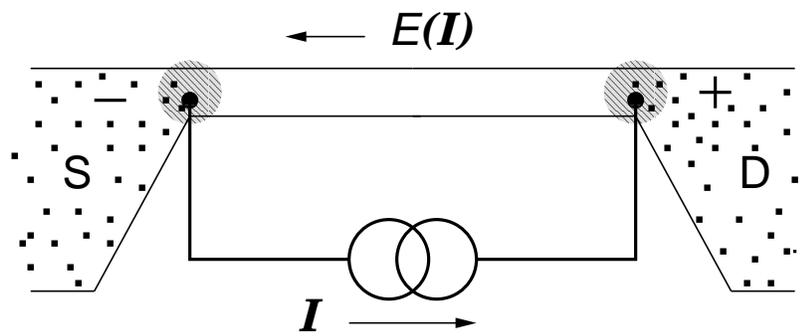
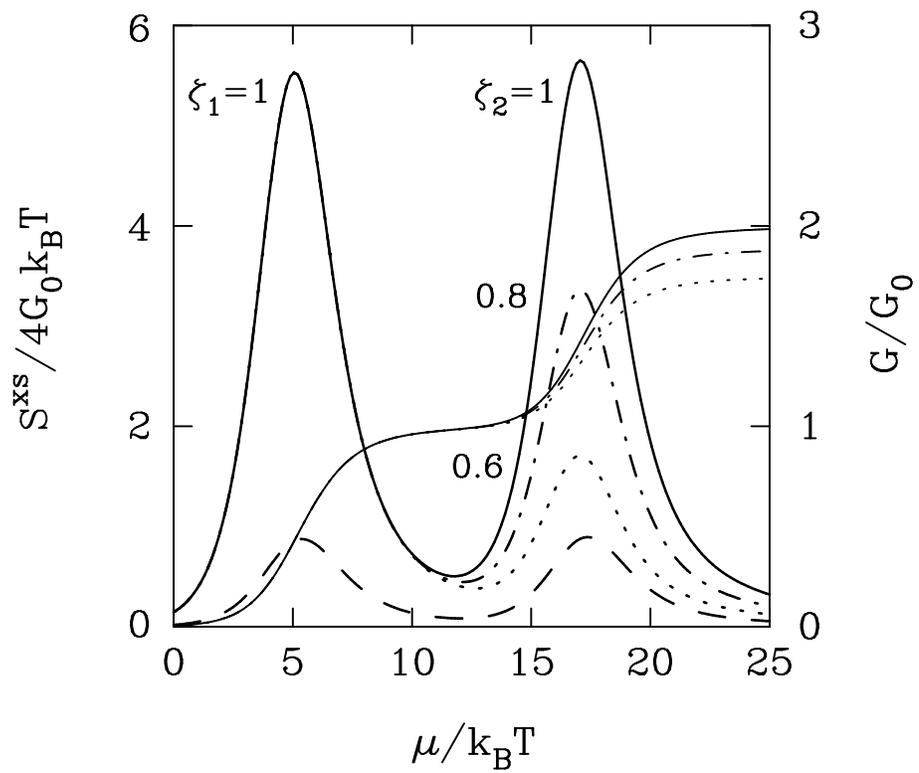
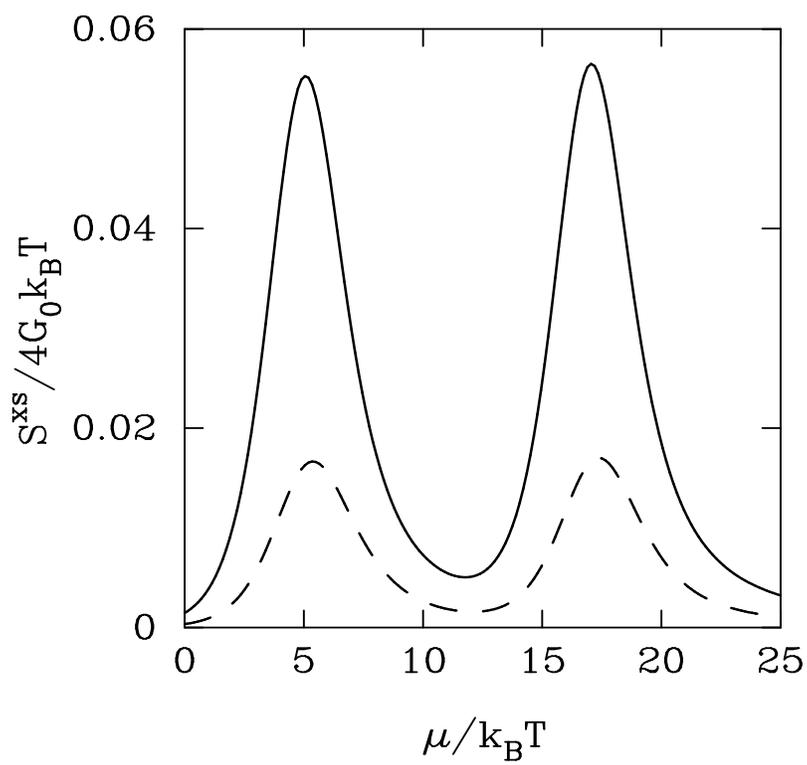
\end{document}